\documentclass[final,1p,times,twocolumn,authoryear]{elsarticle}

\usepackage{amssymb}

\journal{New Astronomy}

\begin{document}

\begin{frontmatter}



\title{ROLE OF FEEDBACK IN AGN-HOST COEVOLUTION: A STUDY FROM PARTIALLY OBSCURED ACTIVE GALACTIC NUCLEI}


\author{J. Wang\corref{cor1}}
\cortext[cor]{Corresponding author}
\ead{wj@bao.ac.cn}
\address{National Astronomical Observatories, Chinese Academy of Sciences, 
20A, Datun Road, Chaoyang District, Beijing, China, 100012}

\begin{abstract}
Partially obscured AGNs within a redshift range $z=0.011\sim0.256$ are used to re-study the role of feedback in the AGN-host coevolution issue in terms
of their [OIII]$\lambda$5007 emission line profile.
The spectra of these objects enable us to determine the AGN's accretion properties directly from their
broad H$\alpha$ emission. This is essential for getting rid of the ``circular reasoning'' in our previous study
of narrow emission-line galaxies,
in which the [OIII] emission line was used not only as a proxy of AGN's bolometric luminosity,
but also as a diagnostic of outflow. In addition, the measurement of $D_n(4000)$ index is
improved by removing an underlying AGN's continuum according to the corresponding broad H$\alpha$ emission.
With these improvements, we confirm and
reinforce the correlation between $L/L_{\mathrm{Edd}}$ and stellar population age. More important is that
this correlation is found to be related to both [OIII] line blue asymmetry and bulk blueshift velocity,
which suggests a linkage between SMBH growth and host star formation through the feedback process.
The current sample of partially obscured AGNs shows that the 
composite galaxies have younger host stellar population, higher Eddington ratio, 
less significant [OIII] blue wing and smaller bulk [OIII] line shift than do the Seyfert galaxies .  

\end{abstract}


\end{frontmatter}


\section{INTRODUCTION}
\label{introduction}

Although the nature is still not fully understood at present, active galactic nuclei (AGNs) are
widely believed to co-evolve with their host galaxies (see Kormendy \& Ho 2013 for a recent review).
The concept of co-evolution mainly stems from two observational facts:
1) the tight correlations between the mass of the central supermassive blackhole (SMBH) and several properties of the bulge
of the host galaxy, including the velocity dispersion, luminosity and mass of the bulge (e.g., Magorrian et al. 1998; Gebhardt et al. 2000; 
Merritt \& Ferrarese 2001; McLure \& Dunlop 2002; Tremaine et al. 2002; Haring \& Rix 2004; Ferrarese \& Ford 2005; 
Aller \& Richstone 2007; Gultekin et al. 2009; Woo et al. 2010); 2) both AGN's accretion and star formation 
have a peak at similar redshifts of $z=2\sim3$. (e.g., Ueda et al. 2003; Croom et al. 2004; 
Hasinger et al. 2005; Nandra et al. 2005; Silverman et al. 2008; Shankar et al. 2009;  
Aird et al. 2010; Assef et al. 2011).

What is associated with the gas fall onto a SMBH is the AGN's feedback that results in an interaction between
the energy released in the accretion and the gas in the host galaxy (see a review in Fabian 2012).
In the local universe, a kpc-scale outflow has been frequently identified by spatially resolved
spectroscopy for both ionized (e.g., Holt et al. 2008; Fu \& Stockton 2009; Rupke \& Veilleux 2011;
Westmoquette et al. 2011) and molecular (e.g., Feruglio et al. 2010; Alatalo et al. 2011) gas.
Recent observations point out that the interstellar medium (ISM) throughout the host can be photoionized and kinematically
disturbed by AGN's feedback (e.g., Fu \& Stockton 2009; Greene et al. 2011). 
In fact, involving a feedback is useful for solving the ``over cooling'' problem in the $\Lambda$ cold dark matter ($\Lambda$CDM) galaxy
formation model in which the predicted massive galaxies are much more numerous than the observed ones (e.g., Ciotti \& Ostriker 2007;
Somerville et al. 2008; Hirschmann et al. 2013), and for reproducing the observed
$M-\sigma_*$ relation, luminosity functions of quasars
and normal galaxies (e.g., Haehnelt et al. 1998; Silk \& Rees 1998; Fabian 1999; Kauffmann \& Haehnelt 2000; 
Granato et al. 2004; Springel et al. 2005; Di Matteo et al. 2005, 2007; Croton et al. 2006; Hopkins et al. 2007, 2008;
Khalatyan et al. 2008; Menci et al. 2008; Somerville et al. 2008; Power et al. 2011; Scannapieco et al. 2012).
Dehnen \& King (2013) recently proposed a new scenario of SMBH growth in which the SMBH accretion disk is formed because
of the feedback. The gas that is swept-up by the feedback finally falls towards the SMBH on near-parabolic orbit when
the feedback weakens.

It has been known for a long time that a large fraction of [OIII] doublet show
blue asymmetry and bulk blueshift with respect to the systemic velocity (e.g., Heckman et al. 1981; Veron-Cetty et al. 2001; Zamanov et al. 2002;
Marziani et al. 2003; Aoki et al. 2005; Bian et al. 2005; Boroson 2005; Komossa et al. 2008; Xu \& Komossa 2009;   
Zhang et al. 2013). The observed [OIII] line profile is generally explained by an interaction between NLR clouds and 
wind from central AGNs, which is
supported by the spatially resolved spectroscopic observations of a few nearby Seyfert 2 galaxies, although
the origin of the wind is still an open issue. 
The observations carried out with Hubble Space Telescope indicate a NLR kinematics with a radial outflow acceleration of form $\upsilon=kr$ at
a radius $r<r_t$ and a deceleration of $\upsilon=\upsilon_{\mathrm max}-kr$ beyond the turnover radius $r_t$ (e.g., Fischer et al. 2013 and
references therein).

By using the two dimensionless shape parameters (i.e., skewness and kurtosis)
to quantify the [OIII]$\lambda$5007 line shape deviation from a pure Gaussian function in
narrow emission-line galaxies, Wang et al. (2011) indicate a trend that AGNs with stronger blue
asymmetries tend to be associated with younger stellar populations. The authors additionally argued that the
trend is likely driven by the co-evolution between AGN's Eddington ratio ($L/L_{\mathrm{Edd}}$, where
$L_{\mathrm{Edd}} = 1.26\times10^{38}(M_{\mathrm{BH}}/M_\odot)\ \mathrm{erg\ s^{-1}}$ is the Eddington
luminosity) and host galaxy, when $L(\mathrm{[OIII]})/\sigma_*^4$ is used as a proxy of $L/L_{\mathrm{Edd}}$.
The argument is mainly based upon the correlation between $L(\mathrm{[OIII]})/\sigma_*^4$ and line skewness.
However, it is generally accepted that a combination of a narrow core Gaussian profile
and an additional blueshifted, broad Gaussian component
is required to reproduce an
observed asymmetric [OIII]$\lambda$5007 line profile. There is therefore a circular argument on which
the measured $L(\mathrm{[OIII]})$ contains the contributions from both components.

In order to avoid the ``circular reasoning'' issue, this paper re-studies the evolution of [OIII] line profile
by focusing on partially obscured AGNs that are excluded in our previous study.
Here, the partially obscured AGNs refer to Seyfert 1.8/1.9 galaxies and composite galaxies\footnote{A galaxy is classified 
as a composite one if it is
located in the [OIII]/H$\beta$ versus [NII]/H$\alpha$ diagnostic diagram between the empirical and
theoretical demarcation lines that separate AGNs from star-forming galaxies. See Section 4.1 for the details.} with
a Seyfert 1.8/1.9-like spectrum.
In addition to the [OIII] line profile and
host stellar population, the spectra of these objects allow us to directly estimate $L/L_{\mathrm{Edd}}$
from their broad H$\alpha$ emission. The partially obscured AGNs are excluded in Wang et al. (2011) because
of the contamination from the central AGN's continuum, which generally results in an underestimation of the modeled
stellar population age. In this paper, the underestimation is alleviated by properly removing the underlying AGN's continuum
from each observed spectrum.

The paper is organized as follows. The sample selection and spectral analysis are presented in Section 2 and
Section 3, respectively. The results are shown in Section 4, and
the implications are discussed in Section 5. A $\Lambda$CDM cosmology
with parameters $\mathrm{H_0=70\ km\ s^{-1}\ Mpc^{-1}}$, $\Omega_{\mathrm m}=0.3$, and $\Omega_{\Lambda} = 0.7$
(Spergel et al. 2003) is adopted throughout the paper.

\section{PARTIALLY OBSCURED AGNS FROM MPA/JHU SDSS DR7 CATALOG}
A sub-sample of partially obscured AGNs has already been identified by Wang et al. (2011) from \rm
the value-added SDSS Data Release 7 (Abazajian et al. 2009) Max-Planck Institute for Astrophysics/Johns Hopkins
University (MPA/JHU) catalog (see Heckman \& Kauffmann 2006 for a review). The sample totally contains
229 broad-line Seyfert galaxies/LINERs and broad-line composite galaxies after removing the duplicates. \rm
Their redshifts range from 0.0111 to 0.025. 
We refer the readers to Section 2 and 3.4 in Wang et al. (2011) for the details of the sample selection.
Briefly, the sample
requires: 1) a spectrum has a median signal-to-noise ratio (S/N) per pixel of the whole spectrum
S/N$>$20 (number of entries: 135,912); 2) all the emission lines used in the traditional Baldwin-Phillips-Terlevich (BPT) diagnostic
diagram (e.g., Veilleux \& Osterbrock 1987) are detected with a significance level of at least 3$\sigma$ (number of entries: 39,384).
3) the [OIII]$\lambda$5007 emission line has S/N$>$30 (number of entries: 10,913); 4) the [OIII]
line width has $\sigma_{\mathrm{obs}}>2\sigma_{\mathrm{inst}}$, where $\sigma_{\mathrm{inst}}=65\ \mathrm{km\ s^{-1}}$ is
the instrumental resolution of the SDSS spectroscopic survey (number of entries: 3,478);  5) redshift does not lie within the range from
0.11 to 0.12\footnote{This is useful for excluding the possible fake spectral proﬁle
caused by the poorly subtracted strong sky emission line [OI]$\lambda$5577 at the observer frame.} 
(number of entries after removing the duplicates: 2,616);
6) a broad H$\alpha$ component can be identified in the spectrum by means of its blue
high-velocity wing after the stellar features are removed from the spectrum. The broad component is
automatically selected by the criterion $F_{\mathrm w}/\sigma_c\geq3$, where $F_{\mathrm w}$ is the specific flux of the 
line wing averaged within the wavelength range from 6500 to 6350\AA\ in the rest frame 
and $\sigma_c$ the standard deviation of the continuum flux within
the emission-line-free region ranging from $\lambda$5980 to $\lambda$6020.

\section{SPECTRAL ANALYSIS}

The 1-Dimensional spectra of the partially obscured AGNs are analyzed by the IRAF\footnote{IRAF is distributed by
National Optical Astronomy Observatory, which is operated by the Association of Universities for Research in Astronomy, Inc.,
under cooperative agreement with the National Science Foundation.} package.
The spectral analysis includes Galactic extinction correction, transformation to the rest frame, starlight component removal and
emission line profile measurement. The Galactic extinction is corrected for
each spectrum  by the color excess $E(B-V )$ taken
from the Schlegel, Finkbeiner, and Davies Galactic reddening
map (Schlegel et al. 1998), by assuming an $R_V = 3.1$ extinction
law of the MilkyWay (Cardelli et al. 1989). The spectrum is then \bf shifted to the rest frame using
\rm the redshift provided by the SDSS
pipelines. The reduced spectrum in the rest-frame of a partially obscured AGN SDSS\,J095156.71+023602.0 is 
shown in Figure 1 as an illustration.

\subsection{Stellar Features Removal and Stellar Population}
The continuum of the partially obscured AGNs is dominated by the starlight component emitted
from the host galaxies. We model the
stellar absorption features from each rest-frame spectrum as the sum of
the first seven eigenspectra. The eigenspectra are built through the
principal component analysis (PCA) method (e.g., Francis et al. 1992; Hao et al. 2005;
Wang \& Wei 2008) from the
standard single stellar population spectral library developed by Bruzual \& Charlot (2003).
A Galactic extinction curve with $R_V=3.1$ is adopted in the
modeling to account for the intrinsic extinction due to the host galaxy.
A $\chi^2$ minimization is performed for each of the spectra over
the rest-frame wavelength range from 3700 to 8000\AA, except for the regions with strong emission lines.
The modeling and subtraction of the starlight component is illustrated in Figure 1 for 
SDSS\,J095156.71+023602.0.  

The modeled starlight component is used to measure host stellar population age.
The 4000\AA\ break index $D_\mathrm{n}(4000)$ (Bruzual 1983; Balogh et al. 1999) defined as
\begin{equation}
D_\mathrm{n}(4000) = \frac{\int^{4100}_{4000}f_\lambda d\lambda}{\int^{3950}_{3850}f_\lambda d\lambda}
\end{equation}
is generally used as an excellent mean age indicator of the stellar population of the bulge of a
galaxy until a few Gyr after the onset of a star formation activity (e.g.,  Kauffmann et al. 2003; Heckman et al. 2004;;
Kewley et al. 2006; Kauffmann \& Heckman 2009; Wild et al. 2007, 2010; Wang \& Wei
2008, 2010; Wang et al. 2011, 2013).

The contamination due to the underlying AGN's continuum is an issue and must be considered in the stellar population synthesis
of partially obscured AGNs. The observed broad H$\alpha$ emission demonstrates the existence of an underlying AGN's continuum leaking from the
edge of the torus according to the unified model (e.g., Antonucci 1993).
The leaking powerlaw continuum impacts the spectral shape at the  blue end,
which always results in an underestimate of $D_n(4000)$ index, and consequently a relatively younger stellar population
if the contamination is ignored (e.g., Cid Fernandes \& Terlevich 1995; Storchi-Bergmann et al. 2000; Cid Fernandes et al. 2004;
Wang et al. 2013). Strictly speaking, the best way to remove the underlying
AGN's continuum is to model the observed continuum by a
linear combination of a power-law continuum and the used eigenspectra.
This is, however, a hard task according to our experiments because of the strong degeneracy between the
AGN's continuum and the blue spectra of hot stars. The degeneracy typically results in a significant overestimate of the amplitude
of the powerlaw component (e.g., Cid Fernandes et al. 2004; Wang et al. 2013).

An alternative approach is adopted here to remove the underlying AGN's continuum.
A scaled powerlaw (i.e., $f_\lambda\propto\lambda^{-\alpha}$) is subtracted from each modeled starlight spectrum that is
dereddened by the modeled local extinction in advance. The amplitude of the powerlaw is estimated from
the tight $L_{5100\AA}$-$L_{\mathrm H\alpha}$ relation (Eq. 4).
The relation finally yields a specific flux of the powerlaw at 5100\AA: $f_{5100\AA}\approx10^{-3}\beta f_{\mathrm H\alpha}$,
where $f_{\mathrm{H\alpha}}$ is the measured line flux of broad  H$\alpha$ emission, and
$\beta$ accounts for the effect on the flux due to reddening.

We at first checked the effect of different powerlaw index $\alpha$ on the measured $D_n(4000)$ value.
The variation of powerlaw shape accounts for not only the intrinsic scatter of $\alpha$ around
its typical value of 1.7 (e.g., Vanden Berk et al. 2001 and references therein),
but also the uncertainty due to reddening of AGN's continuum. When $\alpha$ varies from 1.5 to 1.9,
a tiny change $\leq0.02$ can be obtained for $D_n(4000)$, which is significantly smaller than the
dynamical range ($\sim1.0$) of the measured $D_n(4000)$. This result is not unreasonable because the specific flux ratio
between the two powerlaws with $\alpha=1.5$ and 1.9 has a shape $\propto\lambda^{-0.4}$, which is very close to
a flat spectrum within the narrow wavelength ranges defining $D_n(4000)$.

The left panel in Figure 2 compares the $D_\mathrm{n}(4000)$ index measured with the
powerlaw correction ($[\alpha,\beta]=[1.7,1.0]$) with that without the correction.
One can see clearly from the comparison: 1) as expected, an enhancement of $D_\mathrm{n}(4000)$ value
after the correction of powerlaw; 2) the enhancement moderately increases with the measured
$D_\mathrm{n}(4000)$. The increase could be understood as follows.
In a spectrum with larger 4000\AA\ break, the relative flux level contributed by a powerlaw has a
more significant difference between the wavelength ranges blueward and redward of 4000\AA.
The right panel shows the same comparison but in the case of $[\alpha,\beta]=[1.7,2.0]$\footnote{A factor of $\beta=2$ is used by taking
into account of the factor of $f_e\sim0.47$ obtained in Wang \& Wei (2008). The factor $f_e$ represents an upper limit
on the reducing of broad H$\alpha$ lines emission due to reddening.}.
As expected, a stronger AGN's continuum results in a larger $D_n(4000)$ enhancement and a more significant increasing of
the enhancement with $D_n(4000)$ value. Moreover, a larger scatter could be seen from the comparison for a stronger 
AGN's continuum. 

Without a further statement,
the $D_n(4000)$ value measured after the powerlaw correction with $[\alpha,\beta]=[1.7,1.0]$ is used
throughout the subsequent analysis. The effect on the results by different powerlaw level is discussed in Section 5.2.


\subsection{Emission Line Measurement}
After the removal of the starlight component, the emission-line profiles are at first modeled by the SPECFIT task (Kriss 1994)
in the IRAF package in the emission-line isolated spectra for both H$\alpha$ and H$\beta$ regions\footnote{
These spectral modelings are performed for the two reasons. At first, an integrated narrow-line flux
is required for the subsequent examination in the Baldwin-Phillips-Terlevich (BPT, Baldwin et al. 1981; Veilleux \& Osterbrock 1987)
diagrams. Secondly, the broad H$\alpha$ component is
required to be separated from the observed line profile for the estimation of both SMBH accretion properties
and level of the underlying AGN's continuum.}.
Each line is modeled by a linear combination of a set of several Gaussian profiles.
The intensity ratios of the [OIII] and [NII] doublets are fixed to their theoretical values.
The flux of the [OI]$\lambda$6300 emission line is measured through direct integration by the SPLOT task in
the IRAF package.

By following the method in Wang et al. (2011), the two high-order line shape parameters $\xi_3$ (Skewness) and $\xi_4$ (kurtosis)
are used to quantify the line shape deviation from  a pure Gaussian profile (Binney \& Merrifield 1998).
The parameters are defined as $\xi_k=\mu_k/\sigma^k (k=3,4)$, where
\begin{equation}
 \mu_k=\bigg(\frac{c}{\overline{\lambda}}\bigg)^k\frac{\int(\lambda-\overline{\lambda})^kf_\lambda d\lambda}{\int f_\lambda d\lambda}
\end{equation}
is the k-order moment of line and $\sigma$ the second-order moment in units of $\mathrm{km\ s^{-1}}$ (i.e., $k=2$ in Eq. 2).
$f_\lambda$ is the specific flux density of the continuum-subtracted emission line, and $\overline{\lambda}=\int\lambda f_\lambda d\lambda/\int f_\lambda d\lambda$
the line centroid (i.e., the first moment) of the emission line.
A value of $\xi_3>0$ corresponds to a red asymmetry, and $\xi_3<0$ a blue asymmetry. The emission line
with a value of $\xi_4>3$ has a peak profile superposed on a broad base, while the line with $\xi_4<3$ has a “boxy” line profile\footnote{We
refer the readers to Section 3.5 in Wang et al. (2011) for the uncertainties of the skewness and kurtosis parameters of $\Delta\xi_3=0.14$ 
and $\Delta\xi_4=0.20$ that are derived from the duplicate observations. We emphasize that the sample selection in the current study is the 
same as that in Wang et al. (2011), except for the broad H$\alpha$ emission. }.
Figure 3 presents the $\xi_3$ versus $\xi_4$ diagram for the partially obscured AGNs.
The figure shows a sequence starting from the pure Gaussian region (i.e., $\xi_3=0$, $\xi_4=3$) to the upper left corner for
both Seyfert and transition galaxies with broad Balmer emission, which is highly consistent with that in Wang et al. (2011)
for narrow emission-line galaxies.

The [OIII]$\lambda5007$ line bulk relative velocity shift is calculated as
$\Delta\upsilon=c\Delta\lambda/\lambda_0$, where $\lambda_0$ and $\Delta\lambda$ are the
rest-frame wavelength of the [OIII]$\lambda$5007 emission line and the wavelength
shift with respect to the narrow H$\beta$ line, respectively. The narrow H$\beta$ line shows
a very small velocity shift relative to the galaxy rest frame (e.g., Komossa et al. 2008).
$\Delta\upsilon$ is based on the measured line centroids ($\overline{\lambda}$) and on the
wavelengths in a vacuum of both H$\beta$ and [OIII] lines.
The calculated $\Delta\upsilon$ is compared with the $\xi_3$ (the left panel) and $\xi_4$ (the right panel)
parameters in Figure 4. The trends plotted in the two diagrams are very close to that for narrow emission-line galaxies shown in
Wang et al. (2011). A spearman rank-order test yields a correlation between $\Delta\upsilon$ and $\xi_3$ with
a correlation coefficient of $r_s=0.439$ and a probability of null correlation of $P=1.8\times10^{-11}$ corresponding to 
a Z-value of 4.53. One can see from the
left panel that there is a small fraction of the objects that deviate from the correlation at the large blueshift end.
The right panel indicates that these objects are dominated by the ``boxy'', rather than the peaked, [OIII] line profiles, which is
again in agreement with that in Wang et al. (2011). The authors have argued that a “boxy” line profile could
be reproduced by the sum of two or more
(distinct) peaks with comparable fluxes and line widths, which results in a shift of line centroid.

\subsection{Deriving Physical Properties of Accretion}

The two basic parameters (i.e., $L/L_{\mathrm{Edd}}$ and $M_{\mathrm{BH}}$, where $L_{\mathrm{Edd}}=1.26\times10^{38}M_{\mathrm{BH}}/M_{\odot}$) 
of SMBH accretion are estimated from
the AGN's broad H$\alpha$ line emission that is obtained through our spectral profile modeling.
Greene \& Ho (2007) provided an updated estimator of the mass of central SMBH
\begin{equation}
 M_{\mathrm{BH}}=(3.0^{+0.6}_{-0.5})\times10^6\bigg(\frac{L_{\mathrm{H\alpha}}}{10^{42}\ \mathrm{ergs\ s^{-1}}}\bigg)^{0.45\pm0.03}
\bigg[\frac{\mathrm{FWHM(H\alpha)}}{10^3\ \mathrm{km\ s^{-1}}}\bigg]^{2.06\pm0.06}M_\odot
\end{equation}
by combining the revised luminosity-radius relation (with a scatter of 30-50\% of the measured radius)
reported in Bentz et al. (2006) and the $L_{5100\AA}$-$L_{\mathrm H\alpha}$
relation in Greene \& Ho (2005)
\begin{equation}
 L_{5100\AA}=2.4\times10^{43}\bigg(\frac{L_{\mathrm{H\alpha}}}{10^{42}\ \mathrm{ergs\ s^{-1}}}\bigg)^{0.86}\ \mathrm{ergs\ s^{-1}}
\end{equation}
where $L_{\mathrm{H\alpha}}$ is the intrinsic luminosity of the H$\alpha$ broad component
corrected for local extinction. The luminosity relation has a 
rms scatter around the best-fit line of 0.2dex. The extinction is obtained from the narrow-line ratio H$\alpha$/H$\beta$ for each object, assuming
the Balmer decrement for standard case B recombination and a Galactic extinction curve with $R_V=3.1$. The 
AGN's bolometric luminosity is then derived from the calibration $L=9\lambda L_{\lambda}(5100\AA)$ (Kaspi et al. 2000).

\section{ANALYSIS AND RESULTS}

This section presents the statistical study on the partially obscured AGNs based on
the spectral measurements described above.
The spectral properties of these objects enable us to determine the AGN's accretion properties directly from
their broad H$\alpha$ emission instead of from other proxy. This
is essential for getting rid of the ``circular reasoning'' in our previous study of narrow emission-line galaxies (Wang et al. 2011).
In that study, both bolometric luminosity of AGN and line profile parameters are derived from [OIII]$\lambda$5007 emission line.
In total, 15 objects 
are excluded from the subsequent statistical analysis either because of their poor starlight removal
or because of their poor S/N of the broad H$\alpha$ emission. 

\subsection{BPT Diagnostic Diagrams}

Figure 5 displays the three BPT diagrams for
the final 170 broad-line Seyfert galaxies and
44 broad-line transition galaxies\footnote{Among these 214 partially obscured AGNs, there are 29 common objects listed in the 
\it ROSAT\rm-SDSS-DR5 catalog that is originally crossmatched by Anderson et al. (2007). Their rest-frame X-ray 
luminosities in the 0.1-2.4 keV band range from $10^{42}$ to $10^{44}\mathrm{erg\ s^{-1}}$, \bf  
This range indicates that some of these objects are AGNs (e.g., those having a luminosity above $10^{43}\mathrm{erg\ s^{-1}}$).\rm}. 
All the line ratios are calculated from the narrow Balmer line emission that
is obtained through our profile modeling. 
The [OIII]$\lambda$5007 line flux used in the line ratio calculation
is the sum of both narrow and broad (mostly blueshifted) components 
because these
two components are required to properly model the observed [OIII] line profile.     
The diagrams are a powerful tool in determining
the dominant powering source in narrow emission-line galaxies through their emission-line ratios.
The solid lines in the three panels show the theoretical demarcation lines separating ``pure'' AGNs from star-forming galaxies
(Kewley et al. 2001). The long-dashed and dotted line in the left panel mark the
empirical demarcation line proposed in Kauffmann et al. (2003) and the theoretical one proposed
by Stasinska et al. (2006), respectively. Based upon their new photoionization models, Stasinska et al. (2006)
claimed that their demarcation line is slightly restrictive
in separating ``pure'' starforming galaxies than the empirical one.
As shown by the diagrams, almost all the objects listed in the sample are located above these demarcation lines, and our sample is
strongly biased against LINERs that are typical of low [OIII]/H$\beta$ ratio. The bias is caused by our sample selection that
requires a high S/N ratio (and consequently a large flux) of [OIII] emission line.

\subsection{[OIII] Line Profile Versus Stellar Population}

The measured $D_n(4000)$ value is plotted against $\xi_3$ (the left panel) and
$\Delta\upsilon$ (the right panel) in Figure 6. 
At first, the diagrams confirm the
results obtained in the previous studies (e.g.,Kewley et al. 2006; Schawinski et al. 2007; Wang \& Wei 2008; Wang et al. 2011):
the transition galaxies are clustered in the region with younger stellar populations than do Seyfert galaxies, which implies that
the transition galaxies are at an intermediate evolutionary phase in the context of the co-evolution of AGNs and their
host galaxies (see discussion and references in Wang et al. 2011). Secondly, although no evident correlation can be identified from
the two diagrams, a clear difference can be obtained if the sample is separated into two parts according to their stellar
population ages. The separation is marked by a dashed line with $D_n(4000)=1.5$ in each panel. This value is usually used to
define young/old stellar population according to the $D_n(4000)$ measurement (e.g., Kauffmann et al. 2003). The partially
obscured AGNs associated with young stellar populations show a wide range in their [OIII]
line profiles that vary from a blue asymmetrical shape to a Gaussian function. On the contrary, a symmetric
line profile is typical for the AGNs associated with old stellar populations. A similar trend can be also found in the
$\Delta\upsilon$ case. AGNs with significant [OIII] blueshift ($\Delta\upsilon < 150\mathrm{km\ s^{-1}}$)
can be only observed in the galaxies with young stellar populations with $D_n(4000)<1.4$.

\subsection{$L/L_{\mathrm{Edd}}$ Derived From Broad H$\alpha$ Emission}

The left panel in Figure 7 plots $L/L_{\mathrm{Edd}}$ as a function of stellar population age. A significant
correlation can be identified between AGN's accretion power and host galaxy stellar population age: larger the
$L/L_{\mathrm{Edd}}$, younger the host stellar population will be ($r_s=-0.622$, $P=3.1\times10^{-27}$ and $Z=6.72$ for the
Spearman rank-order test). The strong correlation confirms and reinforces the
results that obtained by many authors in the previous studies (see discussion for more details).
 We parametrize the relationship as the best fit of 
\begin{equation}
 \log L/L_{\mathrm{Edd}}=-1.75-2.88x+4.08x^2
\end{equation}
where $x=1/D_n(4000)$, The relationship is over plotted by a dashed line in the diagram.

$L/L_{\mathrm{Edd}}$ is plotted against $\xi_3$ in the middle panel of Figure 7. Similar as the $D_n(4000)$
versus $\xi_3$ diagram, there is no evident relationship between the two parameters at first glance.
However, one can identify a trend that the objects are avoid to occupy the left-bottom corner with
low $L/L_{\mathrm{Edd}}$ and strong [OIII] blue asymmetry. To reveal this trend more clearly, we separate the sample into
two parts basing upon $\xi_3=-0.5$. Figure 8 compares the distributions of $L/L_{\mathrm{Edd}}$ of the two sub-samples.
When compared with the AGNs with weaker [OIII] blue asymmetry, the distribution of the ones with
stronger asymmetry is slightly shifted towards the higher $L/L_{\mathrm{Edd}}$ end.
A two-sided Kolmogorov-Smirnov test yields a marginal difference between the two distributions
at a significance level of $P=0.95$ with a maximum absolute discrepancy of 0.23 in logarithm. 
Mullaney et a. (2013) compares the stacking [OIII] line profiles with various AGN's parameters
for 24,264 optically selected AGNs. Their panel b in Figure 3 indicates a trend in which the strength 
of the [OIII] blue wing increases with Eddington ratio, which is consistent with the result reported here.

The right panel in Figure 7  presents an anti-correlation between $L/L_{\mathrm{Edd}}$ and $\Delta\upsilon$,
which is failed to be identified for narrow emission-line galaxies in Wang et al. (2011).
The correlation indicates that the bulk velocity blueshift of [OIII] emission line increases with
$L/L_{\mathrm{Edd}}$. A Spearman rank-order test returns a correlation coefficient $r_s=-0.319$ at a
significance level of  $P=1.9\times10^{-6}$ ($Z=3.29$). Zhang et al. (2011) recently analyzed a large and
homogeneous sample of radio-quiet Seyfert 1 galaxies and quasars selected from SDSS. Their analysis indicates that
the bulk velocity shift of [OIII] is found to be strongly related with $L/L_{\mathrm{Edd}}$: the large velocity
shift generally occurs in the AGNs with high $L/L_{\mathrm{Edd}}$ (see also in e.g., Boroson 2005; Bian et al. 2005).
As a rare population, the ``blue outliers'' are the objects
with strong [OIII] blueshifts larger than $250\ \mathrm{km\ s^{-1}}$ (e.g., Zamanov et al. 2002; Zhou et al. 2006).
Spectroscopic observations point out that these objects occupy the high $L/L_{\mathrm {Edd}}$ end in the Eigenvector-I (EI) space, and
are exclusively Population A objects\footnote{AGNs are separated into \bf two populations, A (having $\mathrm{FWHM(H\beta)}<4000\ \mathrm{km\ s^{-1}}$ and
large RFe) and B (having $\mathrm{FWHM(H\beta)}>4000\ \mathrm{km\ s^{-1}}$ and small RFe), \rm
where RFe is defined as the FeII to broad H$\beta$ line ratio. 
Generally speaking, \bf Population A AGNs have higher $L/L_{\mathrm{Edd}}$  than do Population B AGNs (see
citations in the main text). \rm Population A
AGNs contain Narrow-line Seyfert 1 galaxies that are believed to \bf be \rm at early evolution stage (e.g., Mathur 2000).   
} 
that have small broad H$\beta$ line widths
($<4000\ \mathrm{km\ s^{-1}}$) and strong FeII complex emission (e.g., Zamanov et al. 2002; Marziani et al. 2003; Komossa et al. 2008;
Marziani \& Sulentic 2012).

Compared with the Seyfert galaxies, one can learn from Figure 7 that the composite galaxies in the current sample have 
extreme properties with younger host stellar population, higher Eddington ratio, less significant [OIII] blue wing 
and smaller bulk [OIII] line shift.   

\section{DISCUSSION}

\subsection{An Underestimation and Bias of $L/L_{\mathrm{Edd}}$}

As discussed in Wang \& Wei (2008), although $L_{\mathrm{H\alpha}}$ is an acceptable measurement of the accretion power
in partially obscured AGNs, a systematical underestimate of $L/L_{\mathrm{Edd}}$ and $M_{\mathrm{BH}}$ can not be
avoided both because of the obscuration of the torus and because of the reddening in the broad-line region.
Our previous study obtained an upper limit of underestimate of $\sim50\%$ for $L/L_{\mathrm{Edd}}$ and $\sim70\%$ for $M_{\mathrm{BH}}$.

The underestimate of $L_{\mathrm{H\alpha}}$ could be learned in the left panel of Figure 9,
which compares the distribution of $L_{\mathrm{H\alpha}}$ between the used partially obscured AGNs and typical
type I AGNs in Greene \& Ho (2007)\footnote{\bf The adopted bolometric correction coefficient is 9 in the current study, and
is 9.8 in Greene \& Ho (2007).\rm }. The difference between the peaks of the two distributions is about 0.5\,dex.
This difference is likely caused by the reddening and obscuration of the broad H$\alpha$ emission. The middle panel
compares the calculated $M_{\mathrm{BH}}$ between the two samples. A strong bias against small $M_{\mathrm{BH}}$ could be
identified in the comparison. The bias is caused by our broad H$\alpha$ selection. The selection is based on the high velocity
H$\alpha$ line wing over the wavelength range from 6500 to 6530\AA\ (i.e., blueward of [NII]$\lambda6548$
line) in the rest frame, which naturally results in a significant loss of the objects with small
$\mathrm{FWHM_{H\alpha}}\lesssim2000\mathrm{km\ s^{-1}}$. A similar comparison is shown in the right panel for
the calculated $L/L_{\mathrm{Edd}}$. Compared with the type I AGNs, the $L/L_{\mathrm{Edd}}$ distribution of the partially
obscured AGNs is systematically shifted to low $L/L_{\mathrm{Edd}}$ end by an amount of $\sim$1.0\,dex, although both
distributions have roughly identical dynamical range of $\sim$2.0\,dex. This systematical shift could be explained by
the combination of the
underestimate of $L_{\mathrm{H\alpha}}$ and the selection bias of broad H$\alpha$ line width.

\subsection{Influence of Removal of AGN's Continuum}

Comparing with our previous studies, the current measurement of $D_n(4000)$ index is improved by subtracting
an underlying AGN's continuum from the observed spectrum according to the broad H$\alpha$ emission.
As stated in Section 3.1, there is an uncertainty for the moved powerlaw continuum
due to the uncertainties of reddening and obscuration and the intrinsic scatter of the powerlaw index. We have demonstrated
that the measured $D_n(4000)$ is more sensitive to reddening factor $\beta$ than to powerlaw index.
A Monte-Carlo simulation with 1,000 random experiments is then performed to quantitatively examine
the effect on the $L/L_{\mathrm{Edd}}$-$D_n(4000)$ correlation caused by the uncertainty of removed powerlaw continuum level.
For each object, a random $D_n(4000)$ is produced by
a random sampling of $\beta$ in the range from 1 to 2. A Spearman rank-order test is then performed for each of the
1,000 random samples. Figure 10 displays the distribution of the simulated correlation coefficient $r_s$.
The distribution has a mean value of $\overline r_s=-0.604$ and a standard deviation of $5.0\times10^{-3}$.
the simulation indicates that the relationship between $L/L_{\mathrm{Edd}}$ and $D_n(4000)$ is not sensitive to the
uncertainty of the amplitude of the used powerlaw\footnote{In the 1,000 experiments, all the calculated probabilities that the two parameters are
not correlated are $\sim10^{-27}$.}, although the relationship can be slightly degraded by the
uncertainty of the level of the removed powerlaw continuum.

\subsection{Feedback in Co-evolution of AGNs and Their Host Galaxies}

The relationship between $L/L_{\mathrm{Edd}}$ and stellar population age of host galaxy has been
established by different approaches in past decade. This relationship allows us to
establish a co-evolution scenario in which AGNs evolve from a high-$L/L_{\mathrm{Edd}}$ state
to a low-$L/L_{\mathrm{Edd}}$ state as the circumnuclear stellar population continually ages.
In a wide range of  SMBH masses, the majority of local SMBH growth is found to occur not only
in the high accretion phase but also in the galaxies with young stellar
populations and with ongoing or recent star formations (Goulding et al. 2010; Heckman \& Kauffmann 2006 and
references therein). Wang et al. (2006) reported a relation between the EI space and
middle-far-infrared color $\alpha(60,25)$ by performing a PCA analysis on a sample of \it IRAS\rm-selected Seyfert 1.5 galaxies.
This relation naturally indicates a relation between $L/L_{\mathrm{Edd}}$ and host galaxy stellar population,
both because the EI space is commonly accepted to be driven by $L/L_{\mathrm{Edd}}$ (e.g., Boroson 2002) and
because the color $\alpha(60,25)$ addresses the relative importance of AGN activity and starburst activity.
A direct correlation between $L/L_{\mathrm{Edd}}$ and host stellar population age is established either by
using $L(\mathrm{[OIII]})/\sigma_*^4$ as a proxy of $L/L_{\mathrm{Edd}}$ in type II AGNs (e.g., Kewley et al. 2006; Wild et al. 2007;
Kauffmann et al. 2007) or by estimating $L/L_{\mathrm{Edd}}$ directly from
the Balmer broad lines in partially obscured AGNs (Wang et al. 2008, 2010, and this paper).
The role of $L/L_{\mathrm{Edd}}$ in the coevolution issue is
additionally revealed from AGN's hard X-ray emission,
basing upon an identified correlation between AGN's hard X-ray spectral index and
stellar population age of the host galaxy (Wang et al. 2013).

The evolutionary scenario mentioned above describes how AGNs (strictly speaking, which parameters of AGNs) co-evolve with
their host galaxies. This phenomenological scenario, however, does not involve the background physical process that produces
a self-regulated growth of SMBH mass and host galaxy star formation.
One can see from our statistical study that the feedback is a potential ``bridge'' that links and regulates
the growth of SMBH and host galaxy star formation.
According to the aforementioned results, a strong AGN-driven feedback is required for an AGN in its early gas-rich phase
with both rapid SMBH growth (i.e., high $L/L_{\mathrm{Edd}}$) and young host stellar population,
although the physical process that drives the outflow is still an open question. 
In contrast, a weak feedback is sufficient to regulate SMBH mass growth and host star formation in the late gas-poor phase.
The available models include AGN's wind (e.g.,
Crenshaw et al. 2003;  Pounds et al. 2003; Ganguly et al. 2007; Reeves et al. 2009; Dunn et al. 2010; Tombesi et al. 2012),
radiation pressure (e.g., Alexander et al. 2010) and radio jets (e.g., Rosario et al. 2010). 
In the current sample, the non-detection of the connection between radio emission and [OIII] line profile (see Section 5.4) allows 
us to believe that the radiation pressure is probably the initial driver of the outflows.
Matsuoka (2012) recently identified a
deficit of extended emission-line region in the AGNs with high $L/L_{\mathrm{Edd}}$. The deficit could
be explained by the AGN's feedback that blows the gas around SMBH away.
Moreover, the spectroscopic observations using integral-field unit suggest the similar
signatures of outflow in high-redshift galaxies/quasars (e.g., Nesvadba et al. 2008;
Cano-Diaz et al. 2012; Harrison et al. 2012), where the outflow from AGN is predicted to be
much stronger than in local universe according to the model of cosmic galaxy formation (see a more 
extended discussion on this issue at the end of this subsection).

The revealed dependence between feedback and AGN-host co-evolution provides an evidence that supports
the hypothesis that the host star formation is regulated by SMBH growth through the information
provided by AGN-driven feedback. Both AGN-suppressed and AGN-induced host star formation have been
proposed in past decades,  although we can not say which one occurs in AGNs based on our current results.
A commonly accepted mechanism that the feedback works is that
the feedback from central AGN sweeps out circumnuclear gas, which suppresses star formation and
results in a self-regulated SMBH growth and host star formation in both galaxies merger and secular
evolution scenarios (e.g., Alexander \& Hickox 2012; Fabian 2012; Kormendy \& Ho 2012;).
In the color versus absolute magnitude diagram,
there is an excess of AGNs in the ``green valley'' between the blue cloud and red sequence
(e.g.,  Sanchez et al. 2004; Nandra et al. 2007; Schawinski et al. 2009; Treister et al. 2009;
Georgakakis et al. 2009; Cisternas et al. 2011; Mullaney et al. 2012; Rosario et al. 2012).
The excess implies that AGNs are at a transition from the blue cloud to the red sequence, which is
likely resulted from a suppression of star formation due to the feedback from
AGNs\footnote{Recent direct measurements of
star formation through [OII] emission line and infrared emission indicate that these AGNs are in 
reddened star-forming galaxies, rather than in a transition phase (e.g., Silverman et al. 2009; 
Mullaney et al. 2012; Rosario et al. 2013). Xue et al. (2010) indicates that AGN hosts and non-AGN galaxies occupy the same region in
the color-magnitude diagram by using mass-matched samples.} (e.g., Nandra
et al. 2007; Hickox et al. 2009; Hopkins \& Elvis 2010).
Zubovas et al. (2013 and references therein) recently performed a study on positive feedback through
well-resolved numerical simulations of AGN's feedback process.
At early gas-rich phase, the energy-driven feedback compresses the cold gas in host galaxy into dense shells and clumps,
which triggers star formation if the shocked gas cools rapidly. A negative feedback that quenches the star formation,
however, occurs at late gas-poor phase.

The model in King (2003, 2005) indicates that the $M_{\mathrm{BH}}-\sigma_\star$ relation is established through the 
momentum-driven flow with a mildly relativistic velocity of $\upsilon_{\mathrm w}\sim\eta c$, where $\eta\sim0.1$ is the accretion 
efficiency. However, a energy-driven outflow at large scale is expected after the $M_{\mathrm{BH}}-\sigma_\star$ relationships established.
The outflow velocity is $\upsilon_{\mathrm{out}}\simeq925\sigma_{200}^{2/3}(\lambda_{\mathrm{Edd}}f_c/f_g)^{1/3}\mathrm{km\ s^{-1}}$
(King et al. 2011, Zubovas \& King 2012),
where $\sigma_{200}$ is the velocity dispersion in bulge, $\lambda_{\mathrm{Edd}}$ Eddington ratio of central SMBH, $f_c=0.16$ the cosmology
value of the baryon-to-dark-matter density fraction, and $f_g$ the gas fraction in bulge that is of the same order of $f_c$.    
In the current sample, we have the averaged values of $\overline{\sigma}_\star\approx100\mathrm{km\ s^{-1}}$ and 
$\overline{\lambda}_{\mathrm{Edd}}\approx0.04$. With these values, the outflow velocity at large scale is expected
to be $\upsilon_{\mathrm{out}}\sim200\mathrm{km\ s^{-1}}$, which is close to the velocity shift measured from the [OIII]
emission lines.

Large-scale outflow is expected to be strong in high redshift ($z\sim2$) AGNs where the peaks of both AGN's activity 
and star formation occur roughly coincident (e.g., Ishibashi et al. 2013). 
Searching for the relationship between the strength of AGN's feedback and 
effect on host properties in high$-z$ universe is important for understanding the impact of the feedback on the galaxy formation
and evolution. In spite of the debates (e.g., Santini et al. 2012; Harrison et al. 2012), Page et al. (2012) claims a 
suppressed star formation in powerful AGNs in the redshift range from $z=1$ to 3 basing upon the Herschel SPIRE observations 
in sub-millimeter. 

\subsection{Radio Emission And [OIII] Line Profile}
The interaction between radio jet and ISM is another possible mechanism that drives outflow in
galactic-scale (e.g., Holt et al. 2008; Nesvadba et al. 2008; Guillard et al. 2012).
 An outflow of cold, neutral gas due to jet-ISM interaction is identified
by many observations (e.g., Morganti et al. 2005, 2007; Holt et al. 2011; Mahony et al. 2013).
There is an alignment between radio jet and structure in (extended) NLR in many objects (e.g.,
Whittle \& Wilson 2004; Rosario et al. 2010).
Strong blue wings of [OIII]$\lambda$5007 lines have been identified by Brotherton (1996) in a sample of
radio-loud quasars. The [OIII] line width is found to be correlated with radio luminosity at 1.4GHz ($L_{\mathrm{1.4GHz}}$) for
flat-spectrum radio galaxies in early studies (e.g., Heckman et al. 1984; Whittle 1985). By using a large sample
of AGNs (both type I and type II) from SDSS, Mullaney et al. (2013) claimed that the [OIII] line width
is more strongly related to $L_{\mathrm{1.4GHz}}$ than the other AGN's parameters.

The partially obscured AGNs listed in the current sample are cross-matched with the FIRST survey catalog (Becker et al. 2003)
to examine the connection between their radio emission and [OIII] line profiles. The cross-match returns 69
Seyfert galaxies and 26 composite galaxies with detected radio flux exceeding the FIRST limiting flux density (5$\sigma$) of
1mJy. Their radio fluxes range from 1.0 to 734mJy.
All these objects have a radio emission with S/N$>3$. An underlying contribution from host starburst is a 
possible explanation of the high fraction of 
radio detection in current sample ($\sim45$\%) than in typical type-I AGNs ($\sim10$\%, e.g., Rafter et al. 2009; Mullaney et al. 2013).
The most radio-luminous starburst galaxies ($z=0.01-0.05$) have a radio luminosity of $\log P_{\mathrm{5GHz}}=22.3-23.4$ (Smith et al. 1998). 
The high detection rate could be alternatively attributed to our sample selection criterion,
which excludes the majority of less luminous AGNs, such as LINERs, with faint [OIII] line emission.  
The luminosity at 1.4GHz is calculated from
$L_{\mathrm{1.4GHz}}=4\pi d_L^2f_{\nu}(1+z)^{-\alpha-1}$, where $f_{\nu}$ is the integrated flux density at 1.4GHz,
$d_L$ is the luminosity distance, and $\alpha=-0.8$ (e.g., Ker et al. 2012) is
the spectral slope defined as $f_\nu\propto\nu^{\alpha}$. $L_{\mathrm{1.4GHz}}$ is plotted against $\xi_3$ and $\Delta\upsilon$
in the left and right panel in Figure 11, respectively. The calculated radio luminosity ranges from $10^{20}$ to
$10^{25}\ \mathrm{W\ Hz^{-1}}$. One can see from the figure that there is no evident connection between the radio emission
and the [OIII] line profile asymmetry and bulk velocity blueshift in the current sample. The strongest
radio emission (i.e., $L_{\mathrm{1.4GHz}}>10^{24}\ \mathrm{W\ Hz^{-1}}$) in fact occurs in a few of objects with
nearly Gaussian profile and small velocity shift with $-100\mathrm{km\ s^{-1}}$ at most.
The lack of numbers in our partially obscured AGN sample (95 vs. 1,988) is a possible explanation for none detection of
the influence by radio emission seen in Mullaney et al. (2013). 

%

\subsection{Comparison with Type II AGNs}

The spectral analysis in this study allows us to compare the [OIII] line profile between the partially obscured Seyfert galaxies
and the Seyfert 2 galaxies that are studied in Wang et al. (2011). The distributions of line width, $\xi_3$ and $\Delta\upsilon$ are
compared in Figure 12.
The three panels show that the partially obscured AGNs tend to have wider [OIII] line, stronger blue asymmetry and
larger bulk velocity blueshift than do the Seyfert 2 galaxies.

These results are in agreement with previous studies, and are easily understood in the context of the AGN's unified model (e.g., Antonucci 1993).
Vaona et al. (2012) performed a comprehensive comparison of narrow emission line between Seyfert 2 galaxies and intermediate-type Seyfert
galaxies. The authors identified a higher percentage of strong [OIII] line blue asymmetry and large line width in
the intermediate-type Seyfert galaxies than in the Seyfert 2 galaxies.
A similar conclusions are recently obtained in a comparison between type I and type II Seyfert galaxies
in Mullancy et al. (2013). The [OIII] line asymmetry and bulk velocity blueshift are widely used to characterize
AGN's outflows that are initially launched from accretion disc or dusty torus (e.g. Heckman et al. 1981;
Barth et al. 2008; Nesvadba et al. 2008; Greene et al. 2011). In the context of the unified model,
the observed properties of the outflow inferred from [OIII] emission depend on the orientation of the AGN's ionization cone with
respect to the line-of-sight of an observer. A more asymmetrical [OIII] line profile and a larger
bulk velocity are expected in partially obscured AGNs than in Seyfert 2 galaxies if the outflow is along the
axis perpendicular to the plane of torus.

A majority difference between the current study and Wang et al. (2011) is the method used to estimate $L/L_{\mathrm{Edd}}$.
$L/L_{\mathrm{Edd}}$ is obtained through [OIII] line luminosity and stellar velocity dispersion in Wang et al. (2011), and 
through broad H$\alpha$ emission in this paper. Figure 13 compares the different methods in the current partially obscured
AGN sample. The left panel shows a tight correlation between the luminosities of the two emission lines.
One can see clearly a large scatter, a systematical offset and different dynamical ranges when 
the Eddington ratios estimated from H$\alpha$ ($L/L_{\mathrm{Edd}}$) and [OIII] lines ($\lambda_{\mathrm{Edd}}$) are compared with each other in  
the middle panel. Our examination indicates that the large scatter and the systematic are mainly caused by the poor 
relationship between the $M_{\mathrm{BH}}$ determined by H$\alpha$ line and stellar velocity dispersion. 

\section{CONCLUSION}
The role of feedback in the AGN-host coevolution issue is re-studied here by focusing on the [OIII] emission line
profile in partially obscured AGNs selected from SDSS DR7.
The broad H$\alpha$ emission lines enable us to 1) directly determine the AGN's accretion properties;
2) subtract an underlying AGN's continuum from the observed spectrum.
With these improvements, our analysis indicates (confirms) that the AGNs associated with young
stellar population tend to be of large $L/L_{\mathrm{Edd}}$, strong [OIII] blue asymmetry and large [OIII] bulk velocity blueshift ,
which implies that the feedback from central AGN plays an important role in linking as self-regulated
SMBH growth and host star formation.

\it Acknowledgments\\
\rm We thank the anonymous referee for his/her helpful suggestions for improving the manuscript. 
This study uses the
SDSS archive data that was created and distributed by the Alfred P. Sloan Foundation.
The study is supported by the National Basic Research Program of China (973-program, grant No. 2014CB845800)
and by the National Science Foundation of China under grant 11473036. 






\clearpage

\begin{figure}
\begin{center}
\includegraphics[height=5cm]{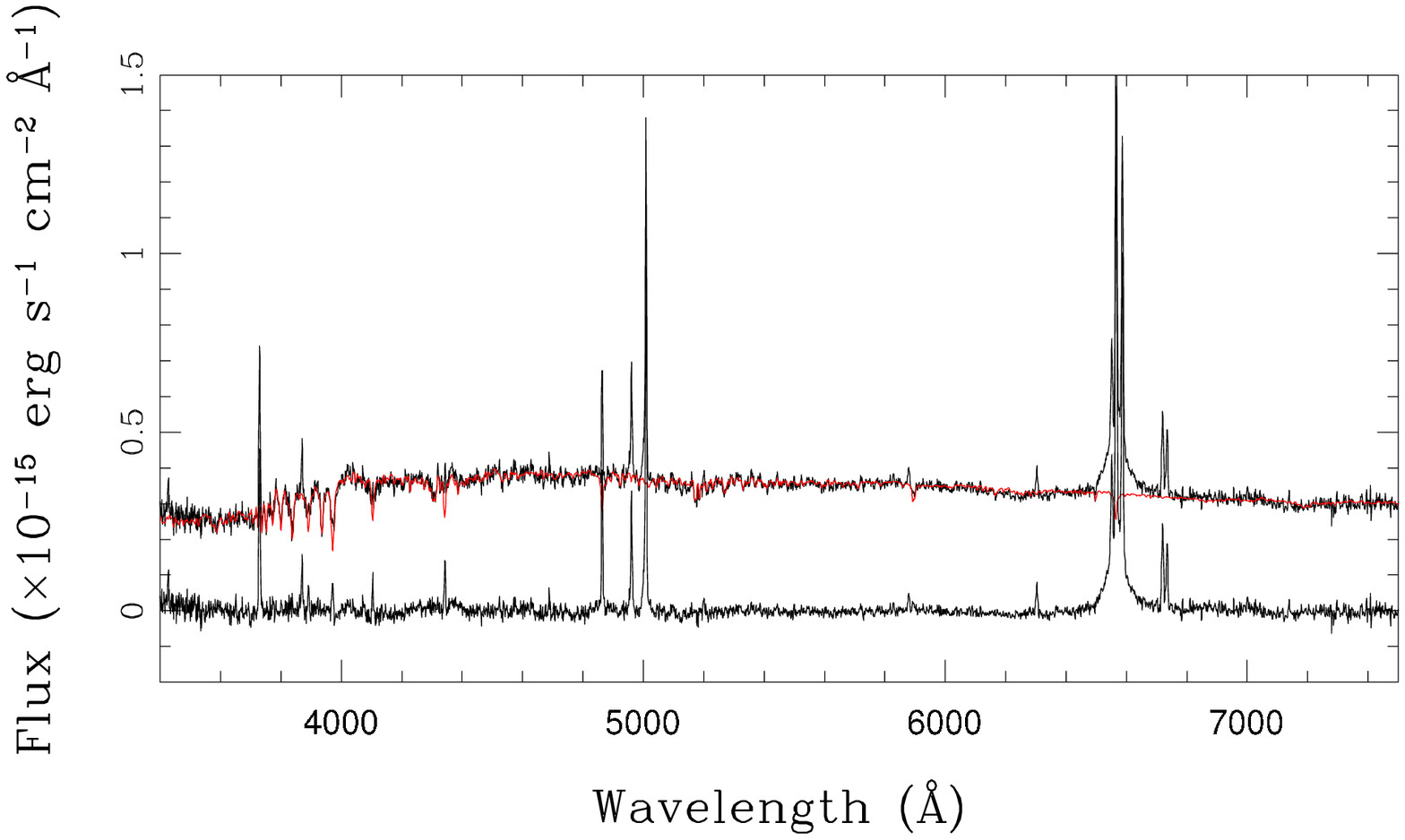}
\caption{An illustration of the continuum modeling and subtraction for a typical case of SDSS\,J095156.71+023602.0. The raw spectrum and emission-line 
isolated spectrum are shown by the two black curves. The red curve presents the modeled continuum by a linear sum of the seven 
eigenspectra built from Bruzual \& Charlot (2003), in which the contamination by the AGN's continuum is not removed. }
\end{center}
\end{figure}

\begin{figure}
\begin{center}
\includegraphics[height=5cm]{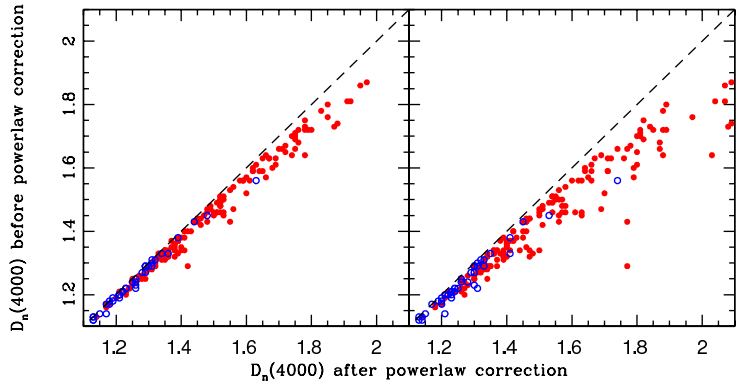}
\caption{A comparison between the $D_n(4000)$ measurements after the correction of an underlying AGN's continuum and the ones
without the correction. \it Left panel: \rm for $\beta=1$; \it Right panel: \rm for $\beta=2$. See text for the
definition of the parameter $\beta$. The Seyfert galaxies and composite galaxies are shown by the red-solid and blue-open circles, respectively.}
\end{center}
\end{figure}

\begin{figure}
\begin{center}
\includegraphics[height=5cm]{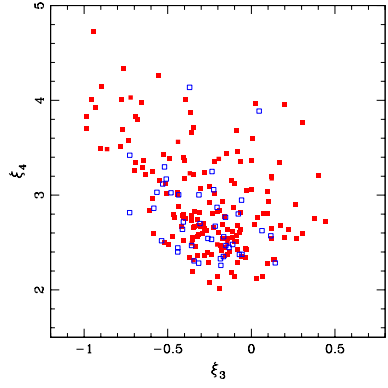}
\caption{The [OIII]$\lambda$5007 line parameter $\xi_4$ (kurtosis) is plotted against $\xi_3$ (skewness) for the partially obscured AGNs.
The Seyfert galaxies and composite galaxies are shown by the red-solid and blue-open squares, respectively.}
\end{center}
\end{figure}

\begin{figure}
\begin{center}
\includegraphics[height=5cm]{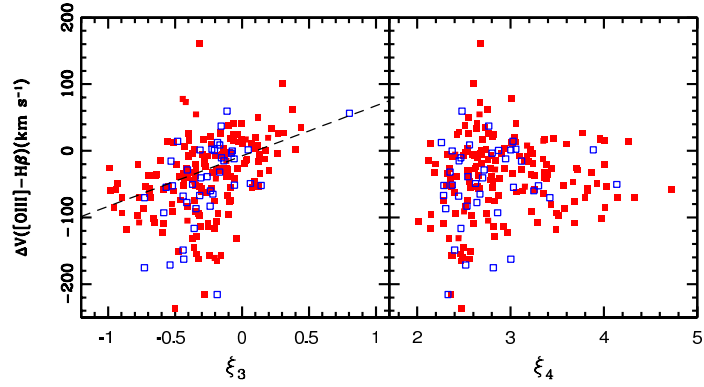}
\caption{Bulk velocity shift $\Delta\upsilon$ of [OIII]$\lambda$5007 line is plotted as a function of the parameters $\xi_3$ (left panel) and $\xi_4$ (right panel)
for the partially obscured AGNs. The symbols are the same as in Figure 3. The dotted line in the left panel shows the best fit of the relationship.}
\end{center}
\end{figure}

\begin{figure}
\begin{center}
\includegraphics[height=5cm]{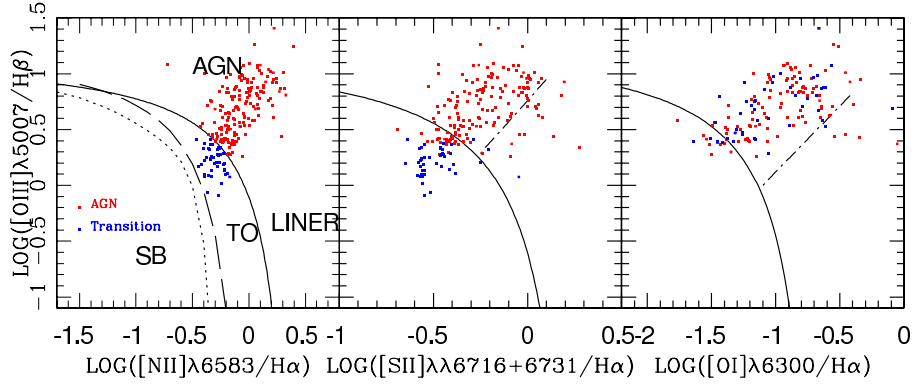}
\caption{Three BPT diagnostic diagrams for the partially obscured AGNs.
The symbols are the same as in Figure 3. The solid lines in all the three panels show the theoretical demarcation lines separating AGNs from star-forming galaxies proposed by
Kewley et al. (2001).  In the left panel, the long-dashed line shows the empirical line proposed in Kauffmann et al. (2003), and the short-dashed
line the theoretical line given in Stasinska et al. (2006). Both lines are used to separate ``pure'' star-forming galaxies. The dashed-dotted lines in
the middle and right panels mark the empirical demarcation lines separating Seyfert galaxies and LINERs proposed in Kewley et al. (2006).}
\end{center}
\end{figure}

\begin{figure}
\begin{center}
\includegraphics[height=5cm]{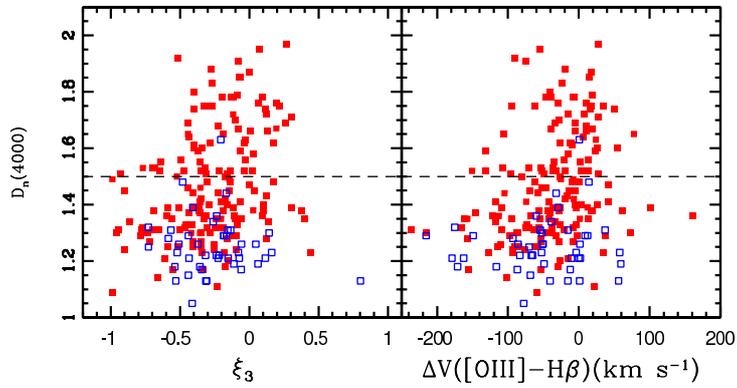}
\caption{Stellar population ages (i.e., the $D_n(4000)$ index) plotted against the [OIII]$\lambda$5007 line shape parameter $\xi_3$ and
bulk velocity shift $\Delta\upsilon$ for the partially obscured AGNs. The symbols are the same as in Figure 3. The horizontal dashed lines in both
panels mark the value of $D_n(4000)=1.5$ that is used to separate the sample into two parts (i.e., with young and old stellar population).
The two sub-samples show a significant difference in the distributions of their [OIII] line shape and bulk velocity shift.
AGNs with old host stellar populations are associated with both symmetric [OIII] line profile and small bulk velocity shift ($>-100\mathrm{km\ s^{-1}}$). 
On the contrary, significant blue wings and bulk blueshift velocities can be only identified in the AGNs with young stellar populations.}
\end{center}
\end{figure}

\begin{figure}
\begin{center}
\includegraphics[height=5cm]{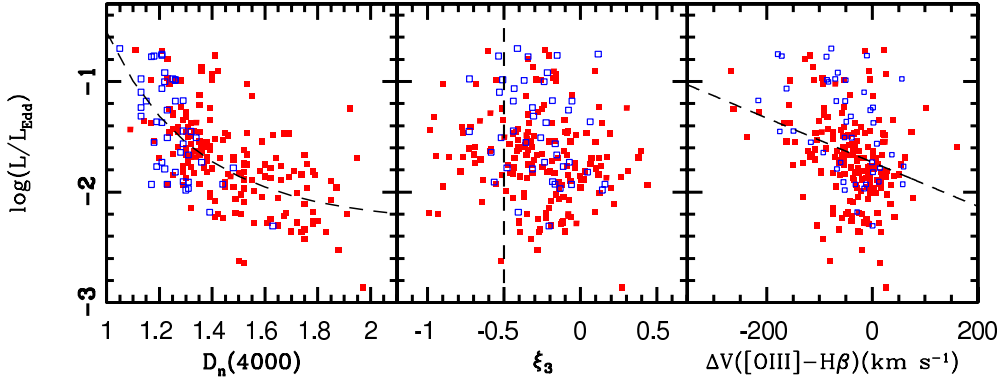}
\caption{\it Left panel: \rm $L/L_{\mathrm{Edd}}$ plotted as a function of stellar population age index $D_n(4000)$. The best fit nonlinear relationship 
(see Eq. 5 in the main text) is over plotted by the dash line. \it Middle and right panels: \rm $L/L_{\mathrm{Edd}}$ 
plotted against $\xi_3$ and $\Delta\upsilon$. The vertical dashed line at $\xi_3=-0.5$ in the middle panel is used to separate the sample into two 
sub-samples (see Figure 8). \bf A best fit by the linear least square regression is shown by the 
dashed line in the right panel. \rm 
The symbols are the same as in Figure 3.}
\end{center}
\end{figure}

\begin{figure}
\begin{center}
\includegraphics[height=5cm]{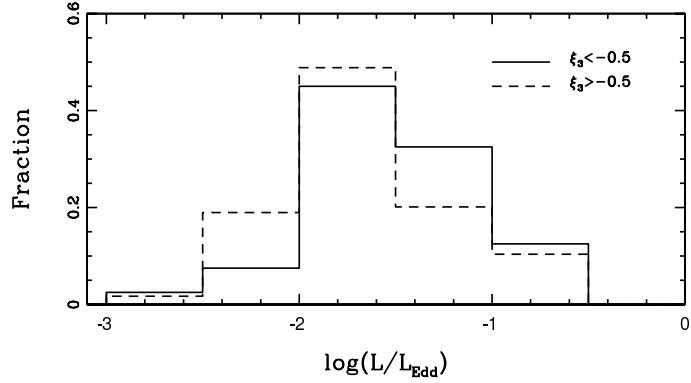}
\caption{Comparison of the distributions of $L/L_{\mathrm{Edd}}$ obtained from broad H$\alpha$ emission line for the two
sub-samples with different [OIII] line shape parameter $\xi_3$. The solid and dashed lines represent the objects
with $\xi_3<-0.5$ and $\xi_3>-0.5$, respectively.}
\end{center}
\end{figure}

\begin{figure}
\begin{center}
\includegraphics[height=4cm]{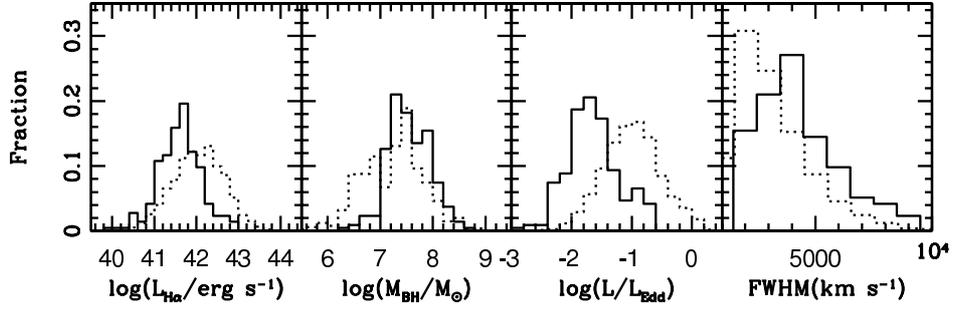}
\caption{A comparison of the distributions of various parameters 
between the used partially obscured AGNs (solid line) and the sample of type I AGNs
(dotted line) in Greene \& Ho (2007). The parameters are luminosity of broad H$\alpha$ emission 
$L_{\mathrm{H\alpha}}$, SMBH mass $M_{\mathrm{BH}}$, Eddington ratio $L/L_{\mathrm{Edd}}$ and 
H$\alpha$ line width from left to right panels. }
\end{center}
\end{figure}

\begin{figure}
\begin{center}
\includegraphics[height=5cm]{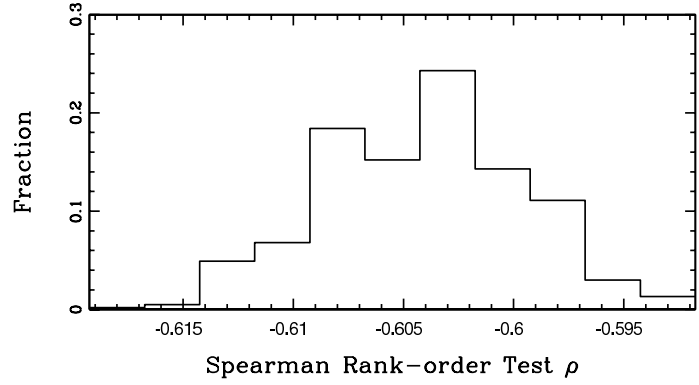}
\caption{Distribution of the simulated Spearman rank-order test coefficient for the $L/L_{\mathrm{Edd}}-D_n(4000)$ correlation. 
A Monte-Carlo simulation with 1,000 experiments is carried out by a random sampling of the level of removed AGN's continuum.
See text for the details of the simulation.
\rm}
\end{center}
\end{figure}

\begin{figure}
\begin{center}
\includegraphics[height=5cm]{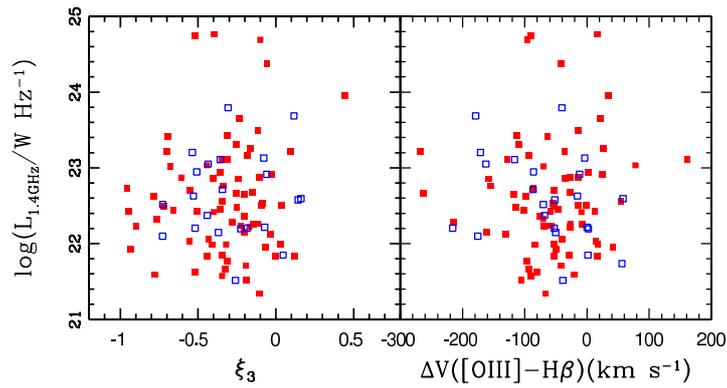}
\caption{The inferred radio luminosity at 1.4GHz plotted against [OIII] line shape parameter $\xi_3$ (left panel) and
bulk velocity shift $\Delta\upsilon$ (right panel). The symbols are the same as in Figure 3.}
\end{center}
\end{figure}


\begin{figure}
\begin{center}
\includegraphics[height=5cm]{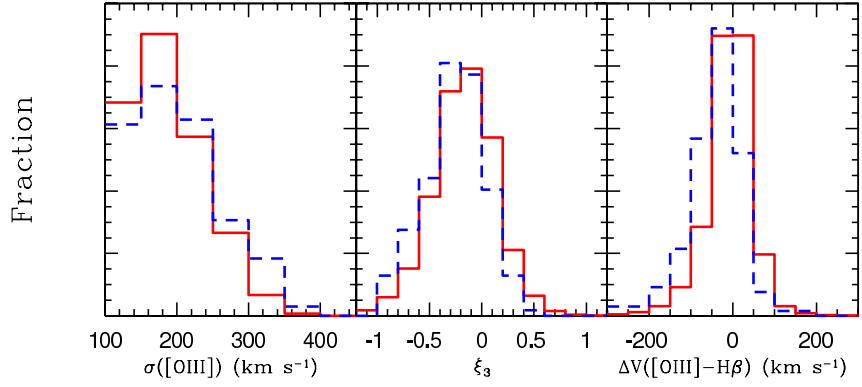}
\caption{\it Left panel\rm: the distribution of [OIII]$\lambda$5007 line width is compared between the partially obscured Seyfert galaxies
(blue-dashed line, this paper) and the Seyfert 2 galaxies (red-solid line) studied in Wang et al. (2011). \it Middle and right panels\rm:
the same as the left panel but for $\xi_3$ and $\Delta\upsilon$, respectively.}
\end{center}
\end{figure}

\begin{figure}
\begin{center}
\includegraphics[height=5cm]{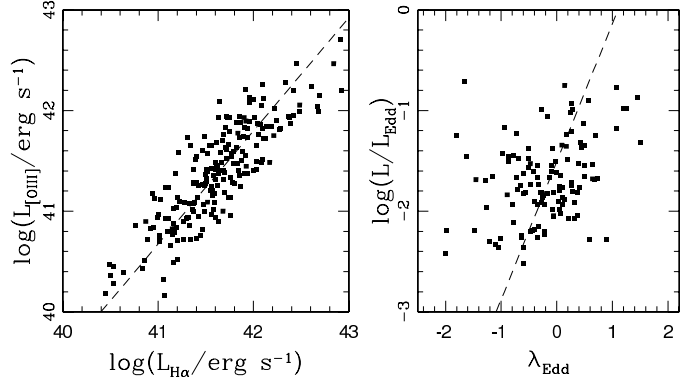}
\caption{\it Left panel\rm: $L_{\mathrm{H\alpha}}$ versus $L_{\mathrm{[OIII]}}$ for the partially obscured AGNs used in the
current study. \it Right panel\rm: Y-axis denotes the $L/L_{\mathrm{Edd}}$ in logarithm estimated from broad H$\alpha$, and 
\bf X-axis \rm the Eddington ratio $\lambda_{\mathrm{Edd}}=\log L/L_{\mathrm{Edd}}$ estimated 
from [OIII] luminosity and stellar velocity dispersion.}
\end{center}
\end{figure}

\end{document}